\documentclass[runningheads,11pt]{llncs}
\pdfoutput=1 
\usepackage[a4paper, total={6in, 8in}]{geometry}
\usepackage{graphicx}
\usepackage{amsmath,amssymb,amstext,times}
\usepackage{color}
\usepackage{float}
\usepackage{booktabs}
\usepackage{hyperref}
\usepackage{multirow}
\usepackage{textcomp}
\usepackage{gensymb}           
\usepackage[symbol]{footmisc}

\usepackage{comment}
\usepackage{algorithm,algorithmic}
\makeatletter
\newcommand{\ALOOP}[1]{\ALC@it\algorithmicloop\ #1%
  \begin{ALC@loop}}
\newcommand{\ENDALOOP}{\end{ALC@loop}\ALC@it\algorithmicendloop}

\makeatother

\newcommand{\bg}[1]{\boldsymbol{#1}}
\newcommand{\bm}[1]{\mathbf{#1}} 

\newcommand\raiseT[2]{%
\setbox0\hbox{$#1{#2}$}\raise\dp0\box0}

\usepackage{todonotes}

\begin{document}
\title{SpectralWeight: a spectral graph wavelet framework for weight prediction of pork cuts}
%
%
\author{
 Majid Masoumi \inst{1,} \thanks{Corresponding Author: majid.masoumi@mail.mcgill.ca.},
 Marcel Marcoux \inst{2},
 Laurence Maignel \inst{3},
 Candido Pomar \inst{2}
}
 
 \authorrunning{M. Masoumi et al.}
 \institute{Montreal Neurological Institute (MNI), McGill University, Montreal, Canada \and
 Sherbrooke Research and Development Centre, Agriculture and Agri-Food Canada, Sherbrooke, Canada \and
 Canadian Centre for Swine Improvement Inc., Ottawa, Canada}
 
%
\maketitle              
\begin{abstract}
In this paper, we propose a novel approach for the quality assessment of pork carcasses using 3D shape analysis. First, we make a 3D model of a pork half-carcass using a 3D scanner and then we take advantage of spectral graph wavelet signature (SGWS) to build a local spectral descriptor. Next, we aggregate the extracted features using the bag-of-geometric-words paradigm to globally represent the half-carcass shape. We then employ partial least-squares regression to predict the weight of pork cuts for the quality assessment of carcasses. Our results demonstrate that SpectralWeight can predict the weight of different pork cuts and tissues with high accuracy. Although in this study we evaluate the performance of SGWS for the weight prediction of pork dissection, our framework is fairly general and enables new ways to estimate the quality and economical value of carcasses of different animals.

\end{abstract}

\section{Introduction}
Quality assessment of hog carcasses has long been practiced in Canada and many other countries \cite{Fredeen:68,Pomar:09}. The quality of a pork carcass can be determined based on its overall body composition by measuring the amount of muscle, fat, skin and bone, or according to the quantity of these tissues inside the primary and commercial cuts. In the literature, there have been different research objectives to evaluate carcasses' quality and cuts. Gispert \textit{et al.} \cite{Gispert:07} characterized pork carcasses based on their genotypes information, and measurements were taken using a ruler and the Fat-O-Meat’er. Marcoux \textit{et al.} \cite{Marcoux:05} employed dual-energy X-ray absorptiometry (DXA) technology to predict carcass composition of three genetic lines with a wide range of varying compositions. Pomar \textit{et al.} \cite{Pomar:03} compared two grading systems based on Destron (DPG) and Hennessy (HGP) probe measurements to verify if both grading approaches result in similar lean yields and grading indices in actual pork carcasses. Engel \textit{et al.} \cite{Engel:03} proposed a different sampling scheme by considering some of the predictive variables to check the accuracy and the approval of new grading systems in slaughterhouses. Picouet \textit{et al.} \cite{Picouet:10} suggested a predictive model based on a density correction equation to determine weight and lean content. In an effort to replace traditional procedures such as dissection, Vester-Christensen \textit{et al.} \cite{Vester-Christensen:09} took advantage of computed tomography (CT)-scans and a contextual Bayesian classification scheme to classify pork carcasses into three types of tissues. The cutout and dissection procedure proposed by Nissen \textit{et al.} \cite{Nissen:06} is a well-recognized reference method to assess the quality of pork carcasses. However, this approach is time-consuming, and financially expensive and requires attention, space and qualified personnel in addition to the risk of bias between butchers \cite{Vester-Christensen:09,Picouet:10}.

Hence, the pork industry, including all stakeholders from production to meat sale, is seeking a way to make the most profitable decisions. One solution is to carry out carcass quality evaluations to know the results coming from a choice of genetic lines, a diet or a breeding method. However, due to the difficulties in conducting the cutting and dissection procedure by butchers, the commercial environment has more constraints than the research environment. Therefore, it becomes more important to develop a simple, fast and precise method to replace the traditional approaches in the commercial environment.

In this paper, we digitize carcasses in three-dimensions using a 3D scanner and then make a triangular mesh model of each pork half-carcass to develop a framework for weight prediction of the different cuts and their tissue composition. Unlike images, triangular meshes have irregular connectivity which demands an efficient and concise design to capture the intrinsic information of the object while staying robust against different triangulation \cite{qiao:19}. This requires the design of a descriptor (signature) that is invariant to isometric deformation of a meshed object while keeping discriminative geometric information \cite{Wang:20}. To this end, we employ a compact signature based on spectral analysis of the Laplace-Beltrami Operator (LBO) to capture the intrinsic geometric properties of shapes. This compact representation of 3D objects simplifies the problem of shape comparison to the problem of signature comparison and provides a relatively accurate prediction of pork cut weights. 

The spectral signatures can be employed in a broad range of applications including medical shape analysis \cite{Masoumi:18b}, 3D object analysis \cite{Bronstein:11,Rodola:SHREC17,Masoumi:17}, shape matching \cite{Melzi:19}, and segmentation \cite{YI:17}. In the literature, there has been a surge of interest in eigenmodes (eigenvalues and eigenvectors) of LBO to build local or global spectral signatures. The power of spectral signatures is mainly due to the spectrum related to the natural frequencies and the associated eigenvectors that yield the wave pattern \cite{Levy:06,Atasoy:16}.

The local spectral signatures are defined on each vertex of a mesh and provide information about the neighborhood around a vertex \cite{Masoumi:19a}. Intuitively, points around a neighborhood share similar geometric information, hence their corresponding local descriptors should represent similar patterns. The local spectral signatures include heat kernel signature (HKS) \cite{Sun:09}, wave kernel signature (WKS) \cite{Aubry:11}, and global point signature (GPS) \cite{Rustamov:07}. From the graph Fourier view, HKS captures information related to the low-frequency component that relies on macroscopic information of a 3D object. Moreover, WKS allows access to the information of the high-frequency component, which corresponds to the microscopic properties of a 3D model. Furthermore, in GPS we might face the problem of eigenvector's switching when their corresponding eigenvalues are close to each other.

On the other side, global signatures encode information about the geometry of the entire 3D object. Shape-DNA~\cite{Reuter:06} was introduced by Reuter \textit{et al.} as a global signature defined by a non-trivial truncated sequence of eigenvalues normalized by mesh area that are arranged in ascending order. Gao \textit{et al.} proposed compact Shape-DNA \cite{Gao:14} by applying the discrete Fourier transform to eigenvalues of the LBO. A new version of GPS developed by Chaudhari \textit{et al.} \cite{Chaudhari:14}, called GPS embedding, is a global descriptor defined as a truncated sequence of inverse square roots eigenvalues of the LBO. However, global spectral signatures give us limited representation and fail to recognize the fine-grained patterns in a 3D model.

Recently, spectral graph wavelet signature (SGWS) has been developed by Masoumi \textit{et al.} \cite{Masoumi:16} as an efficient and informative local spectral signature, which allows analysis of the 3D mesh in different frequencies. Dissimilar to GPS, HKS, and WKS, SGWS leverages the power of the wavelet to provide the information of both macroscopic and microscopic geometry of shape, leading to a more discriminative feature. In this paper, we introduce \textit{SpectralWeight}, in which each 3D model is represented by SGWS to computerize estimation of the weight of pork cuts. Our objective in this study is to verify the accuracy of prediction for different variables of interest and possibly integrate the calculation method into a complete tool that can be used in a commercial environment. To the best of our knowledge, this is the first study on employing SGWS for weight prediction of pork carcasses.

The contribution of this paper is twofold: (1) we propose a framework to precisely model a pork half-carcass by harnessing the power of the spectral graph wavelets, which is called SpectralWeight; and (2) we exploit the SpectralWeight as a predictive model to weigh different cuts of pork.

\section{Material and Methods}\label{Method}

\subsection{Sampling scheme}
\begin{figure}[t]
\setlength{\tabcolsep}{.3em}
\centering
\begin{tabular}{cc}
\includegraphics[scale=.7]{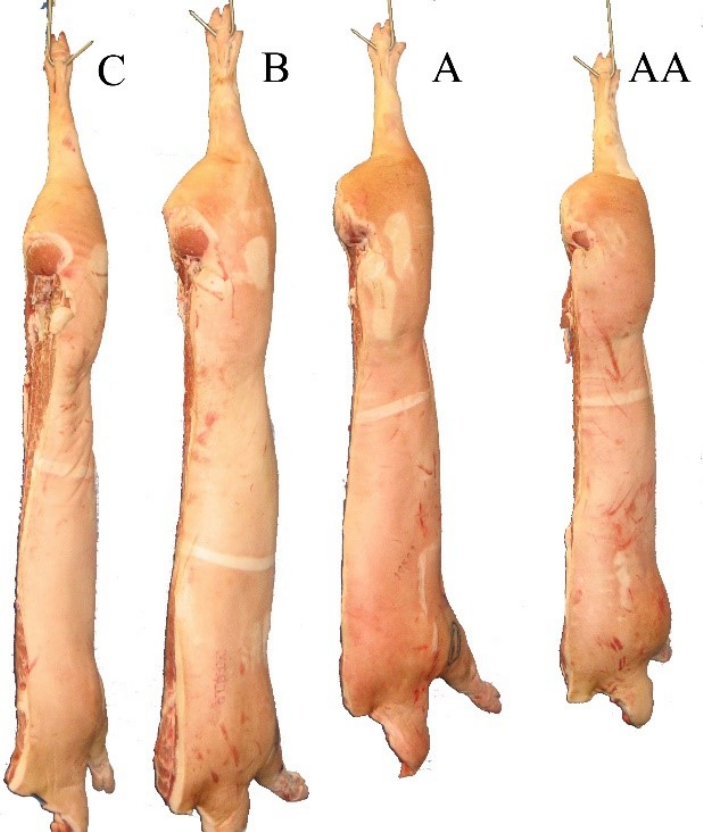}
\end{tabular}
\caption{Pork carcass conformation classes}
\label{Carcass conformation}
\end{figure}

To meet the objectives of this project, we selected $195$ pork carcasses, including $100$ barrows and $95$ females, from commercial slaughterhouses in Quebec, Canada. To obtain a high variability of conformation, carcasses were sampled in a weight range of $83.8$ kg to $116.2$ kg and a backfat thickness range of $7.6$ mm to $30.6$ mm. Backfat thickness was measured using a ruler at the cleft and the level of the fourth-last thoracic vertebra. However, the official backfat measurement was retaken using a digital caliper on a chop cut at the same thoracic level (fourth-last vertebra) $7$ cm from the cleft perpendicular to the skin. The conformation of the carcasses is divided into four different classes represented by the letters C, B, A and AA. Class C represents a long carcass with a thin-looking leg, while class AA represents a stocky carcass with a highly-rounded leg shape (Figure \ref{Carcass conformation}). At the time of weighing, the hot carcass was presented with the head, tail, leaf fat, hanging tender and kidneys. We retained only carcasses properly split in the middle of the spinal column and without tissue ablation. Therefore, each carcass side was considered to be bilaterally symmetric. The scale of variation within each sampling criterion is intended to provide a more robust estimate of the predictive model parameters at the extremes of weight and backfat thickness \cite{Daumas:96}. Only the left half-carcasses were transported to the Sherbrooke Research and Development Center (RDC) of Agriculture and Agri-Food Canada (AAFC). The carcasses were stored in a cooler at $2 \degree C$ in a plastic bag to minimize water loss. The 3D scanning, cutout, dissection and determination of meat cut fat content were completed within days of receipt of the carcasses at the AAFC RDC.

\medskip
\subsection{Half-carcass preparation and modeling}
Before being digitized in 3D, the half-carcasses were prepared in a standard way by removing the tail, the hanging tender and the remains of leaf fat present in the carcass cavity. The jowl was shortened to a uniform length of $15$ cm from the base of the shoulder. The final weight of the half-carcass was subsequently recorded. The total length of the half-carcass was measured using a tape measure from the tip of the rear hooves to the first cervical vertebrae. This length was used to determine the cutting site for the shoulder and ham. Three-dimensional scanning of each half-carcass was performed using the Go!SCAN 3DTM (Model $50$, Creaform, Levis, Quebec, Canada) and post-processed by 3D software (Vxelement, Version 6.3 SR1, Creaform, Levis, Quebec, Canada). The 3D scanner uses white structured light technology without requiring targets affixed to the carcass or additional lighting. Quality control was performed at the beginning of each day by scanning a target provided by the company. All quality controls were passed during the project. The resolution between the mesh points was set to $0.2$ cm and 3D models were saved and used in OBJ format.

\subsection{Cutout and dissection}
\begin{figure}[t]
\setlength{\tabcolsep}{.3em}
\centering
\begin{tabular}{cc}
\includegraphics[scale=1.2]{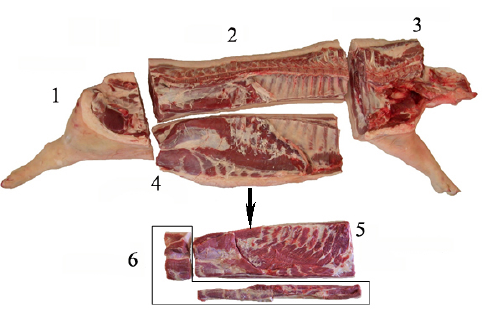}
\end{tabular}
\caption{Four primal cuts, belly commercial trim C400, and belly trimmings (illustrated in the box). (1) primal ham, (2) primal loin (3) primal shoulder (4) primal belly (5) belly commercial trim C400 (6) belly trimmings.}
\label{primal cuts}
\end{figure}

Once the half-carcass was scanned, the four primal cuts (leg, shoulder, loin, and belly) were prepared. The leg and shoulder were cut at proportional distances of $40.90\%$ and $85.54\%$ from the total length of the half-carcass, respectively. These proportions were determined in a previous cutout expertise (unpublished results). The primal loin and flank were separated by applying a straight cut passing $1.5$ cm from the tenderloin and $10$ cm from the base of the ribs opposite the fourth-last thoracic vertebra.
The primal cuts were then prepared into commercial cuts according to different standards. The commercial cuts are presented with or without the skin, more or less defatted, and with or without bone, as appropriate. The skin and ribs from the primal belly were removed. Subsequently, the mammary glands and a portion at the posterior end of the belly (belly trimmings) were cut to create a rectangular appearance. Specifications for the preparation of commercial cuts and their identification codes are presented in the Canadian Pork Handbook and the Distributor Education Program (DEP) \cite{CPI:11}. The cuts illustrated and described in this manual correspond to the basic specifications followed by the Canadian pork industry. It is worth noting that there are no reference numbers for the four primal cuts (Leg, Loin, Shoulder, and Belly) presented in \cite{CPI:11}. Figure \ref{primal cuts}, clearly illustrates the four primal cuts, belly commercial trim $C400$, and belly trimmings. The cutout work resulted in the following parts: Pork leg $C100$, Shoulder blade $C320$, Boneless shoulder blade $C325$, Hock $C355$, Shoulder picnic $C311$, Loin $C200$, Boneless loin $C201$, Skinless tenderloin $C228$, Back ribs $C505$, Belly commercial trim $C400$, Side ribs $C500$, and Belly trimmings. The amounts of bone, skin, and meat (muscle and fat not separated) contained in the primal and commercial cuts were obtained by dissection procedure, and weights were recorded. The meat contained in the main commercial cuts (Pork leg $C100$, Boneless loin $C201$, Loin $C200$, Belly commercial trim $C400$, Shoulder picnic $C311$, Boneless shoulder blade $C325$, Belly trimmings) was minced, and a representative sample was taken to determine lipid, protein and dry matter content using near-infrared transmittance \cite{Shirley:07}. It should be noted that lipid content was used in this study to calculate the weight of fat in the meat of the main commercial cuts. To convert the lipid content to dissected fat weight, a sample of pure muscle and pure fat from each meat mass was also analyzed for lipid content using the same method. Using the data collected from the muscle and fat samples for each cut, an equation was developed to convert the meat lipid content to the dissected fat content. This procedure allows an equivalent amount of fat to be obtained without physically separating the muscle and fat from the meat from the entire mass using a knife.

\subsection{Problem statement and method}
\begin{figure}[t]
\setlength{\tabcolsep}{.3em}
\centering
\begin{tabular}{cc}
\includegraphics[scale=.3]{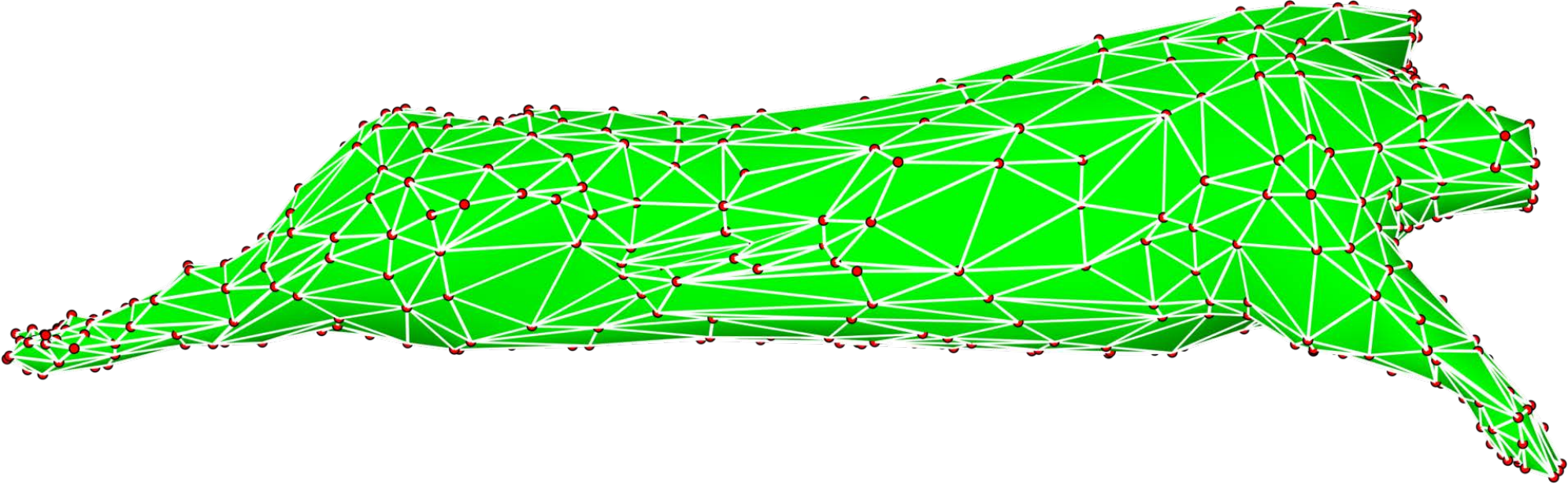}&
 \hspace{1.5cm}
\includegraphics[scale=.3]{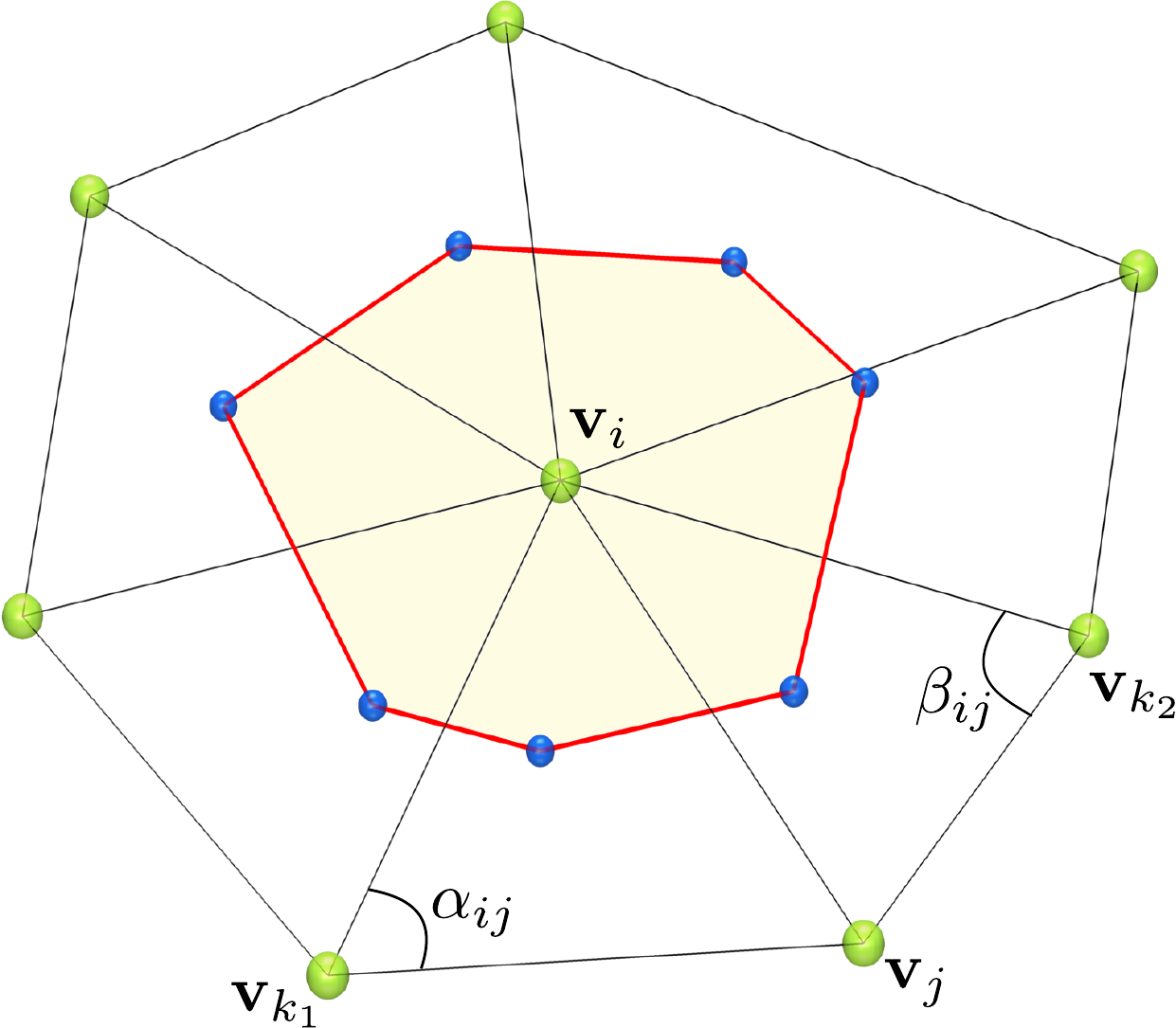}
\end{tabular}
\caption{Triangulated mesh model of a pork half-carcass (Left); illustration of cotangent weight scheme (Right).}
\label{Triangulated carcass}
\end{figure}

We model a pork half-carcass $\mathbb{T}$ as a triangulated mesh defined as $(\mathcal{V},\mathcal{E})$, where $\mathcal{V}=\{\bm{v}_{i}|i=1,\ldots,N\}$ is the set of vertices, and $\mathcal{E}=\{e_{ij}\}$ is the set of edges. For any vertex coordinate $\mathcal{P}=(p_{1},p_{2},p_{3}):\,\mathcal{V}\to\mathbb{R}^{3}$, our objective is to build a local descriptor $f(\bm{v}_{i}) \in \mathbb{R}^{d}$ for each vertex $\bm{v}_{i}$. Figure \ref{Triangulated carcass} (left) represents an example of triangulated mesh on a random pork half-carcass.

\medskip
We build the SpectralWeight framework based on the eigensystem of the LBO that are invariant to the deformation of non-rigid shapes. To achieve the eigenvalues and eigenvectors, we discretize the LBO using a cotangent weight scheme as proposed by \cite{Meyer:03}. We build our Laplacian matrix by:
\begin{equation}
 \bm{L}=\bm{A}^{-1}(\bm{D-W}),
\end{equation}
where $\bm{A}=\mathrm{diag}(a_{i})$ is a mass matrix, $\bm{D}=\mathrm{diag}(d_{i})$ is a degree matrix constructed by $d_{i}=\sum_{j=1}^{n}w_{ij}$, and $\bm{W}=(w_{ij})=\left(\cot\alpha_{ij} + \cot\beta_{ij}\right)/2a_{i}$ is a sparse weight matrix if $\bm{v}_{i}\sim \bm{v}_{j}$. Also, $\alpha_{ij}$ and $\beta_{ij}$ are the angles $\angle(\bm{v}_{i}\bm{v}_{k_1}\bm{v}_{j})$ and $\angle(\bm{v}_{i}\bm{v}_{k_2}\bm{v}_{j})$ of two adjacent triangles $\bm{t}^{\alpha}=\{\bm{v}_{i},\bm{v}_{j},\bm{v}_{k_1}\}$ and $\bm{t}^{\beta}=\{\bm{v}_{i},\bm{v}_{j},\bm{v}_{k_2}\}$, and $a_i$ is the area of the Voronoi cell at vertex $\bm{v}_{i}$, the shaded area. Finally, the eigensystem of LBO is obtained by solving the \textit{generalized eigenvalue problem}, such that:
\begin{equation}
\bm{C}\bg{\xi}_{\ell}=\lambda_{\ell}\bm{A}\bg{\xi}_{\ell},    
\end{equation}
where $\bm{C}=\bm{D-W}$, and $\lambda_{\ell}$ and $\bg{\xi}_{\ell}$ are the eigenvalues and eigenfunctions of LBO, respectively. We define the spectral graph wavelet based for vertex $j$ and scale $t$ as \cite{Masoumi:16}:

\begin{equation}
\bm{s}_{L}(j)=\{W_{\delta_j}(t_k,j)\mid k=1,\dots,L\}\cup\{S_{\delta_j}(j)\},
 \label{Eq:SGWSignatureLevel}
\end{equation}

where $ W_{\delta_j}(t_k,j)$ and $S_{\delta_j}(j)$ are the spectral graph wavelet and scaling function coefficients at resolution level $L$, respectively, as follows (readers are referred to \cite{Masoumi:16} for detailed description):
\begin{equation}
W_{\delta_j}(t,j)=\langle \bg{\delta}_{j},\bg{\psi}_{t,j} \rangle=\sum_{\ell=1}^{m}g(t\lambda_\ell)\xi_{\ell}^{2}(j),
\label{DeltaW_coefficients}
\end{equation}

and

\begin{equation}
S_{\delta_j}(j)=
\langle \bg{\delta}_{j},\bg{\phi}_{t} \rangle=
\sum_{\ell=1}^{m}h(\lambda_\ell)\xi_{\ell}^{2}(j).
 \label{DeltaS_coefficients}
\end{equation}

\begin{figure}[t]
\setlength{\tabcolsep}{.3em}
\centering
\begin{tabular}{ccccc}
\includegraphics[scale=.7]{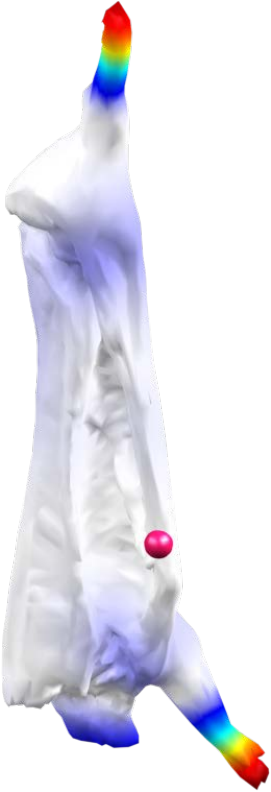}&
\includegraphics[scale=.7]{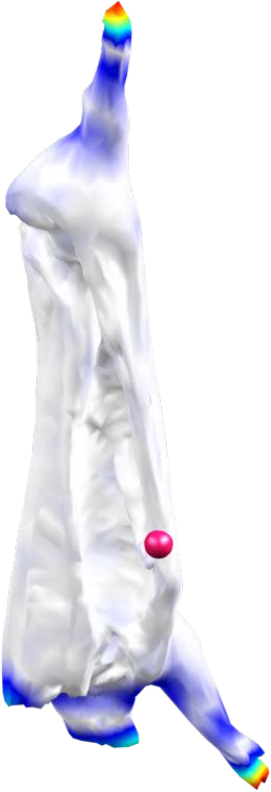}&
\includegraphics[scale=.7]{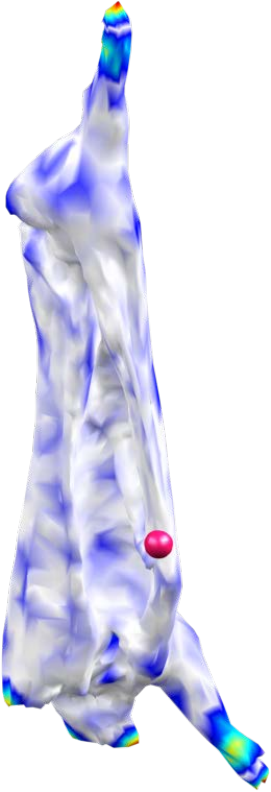}&
\includegraphics[scale=.7]{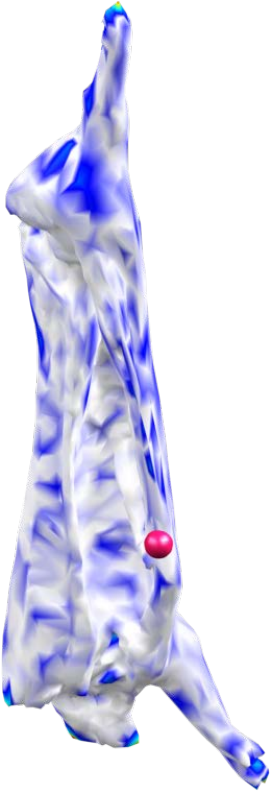}&
\includegraphics[scale=.7]{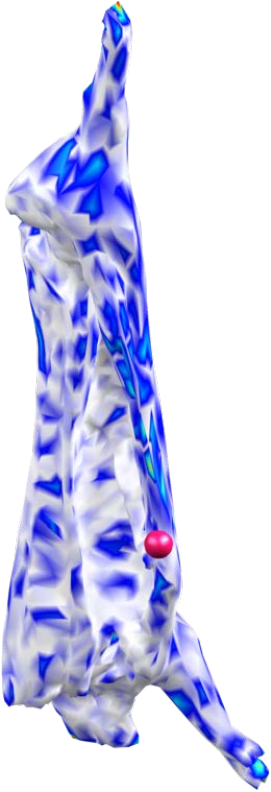}
\end{tabular}
\caption{Visualization of different resolutions (left to right: 1st, 2nd, 3rd, 4th, and 5th) of the spectral graph wavelet signature from a specified point (shown as a pink sphere) on a random pork half-carcass. The cooler and warmer colors represent lower and higher distance values, respectively.}
\label{SGWT representation}
\end{figure}

We consider the Mexican hat wavelet as a generating filter, which treats all frequencies as equally-important and improves the discriminative power of the SpectralWeight. The SpectralWeight takes advantage of nice properties like insensitivity to isometric deformations and efficiency in computation. Moreover, SpectralWeight merges the advantages of both band-pass and low-pass filters for building the local descriptor. Figure \ref{SGWT representation} depicts a representation of SGWT when computing a $\chi^{2}$-distance from a highlighted point on the belly from other points on the carcass. As can be observed, regions with similar geometrical structures share the same color, while regions with dissimilar structures from the specified point bear different colors.

\medskip
The SpectralWeight framework includes the subsequent steps: We first compute an SGWS matrix $\bm{D}$ for each half-carcass in the dataset $\mathcal{S}$, where $\bm{D}=(\bm{d}_{1},\dots,\bm{d}_{m})\in\mathbb{R}^{p\times m}$, and $\bm{d}_i$ is the $p$-dimensional point signature at vertex $i$ and $m$ is the number of mesh points. In the second step, we construct a $p\times k$ dictionary matrix $\bm{V}=(\bm{v}_{1},\dots,\bm{v}_{k})$ through an unsupervised learning algorithm, i.e. clustering, by assigning each $m$ local descriptor into the $k$-th cluster with the nearest mean. In the next step, we employ the soft-assignment coding to map SGWSs $\bm{s}_{i}$ to high-dimensional mid-level feature vectors. This leads to a $k\times m$ matrix $\bm{C}=(\bm{c}_{1},\dots,\bm{c}_{m})$ whose columns are the $k$-dimensional mid-level feature codes. In a bid to aggregate the learned high-dimensional local features, we build a $k \times 1$ histogram $h_{r}=\sum_{i=1}^{m}c_{ri}$ for each half-carcass by sum-pooling the cluster assignment matrix $\bm{C}$. Then, we concatenate the SpectralWeight vectors $\bm{x}_i$ of all $n$ half-carcasses in the dataset $\mathcal{S}$ into a $k\times n$ data matrix $\bm{X}=(\bm{x}_{1},\dots,\bm{x}_{n})$. Afterward, we calculate geodesic distance $g$ \cite{kimmel:98} to extract the diameter of the 3D mesh as well as the volume $v$ of each half-carcass $\mathbb{T}_{i}$ and then aggregate them into $\bm{X}$ to provide further discrimination power for SpectralWeight. Finally, a partial least-squares regression (PLS) is performed on the data matrix $\bm{X}$ to find the equation giving the best fit for a set of data observations. The main steps of SpectralWeight framework are briefly outlined in Algorithm \ref{algo:1}.
\begin{algorithm}
  \caption{SpectralWeight algorithmic steps}\label{algo:1}
  \begin{algorithmic}[1]
    \REQUIRE Set of triangular meshes of $n$ pork half-carcasses $\mathcal{S}=\{\mathbb{T}_1,\dots,\mathbb{T}_n\}$ and their weights $\mathbf{w}$
    \STATE Simplify each model to have a uniform number of vertices.
    \FOR{$j=1$ to $n$}
    \STATE  Compute SGWS matrix $\bm{D}_{j}$ of size $p\times m$ for each half-carcass $\mathbb{T}_{j}$.
    \STATE  Employ soft-assignment coding to determine the $k\times m$ code assignment matrix $\bm{C}_{j}$, where $k>p$.
    \STATE  Represent each half-carcass $\mathbb{T}_{j}$ as a $k \times 1$ histogram $\bm{h}$ by pooling of code assignment matrix $\bm{C}_{j}$.
    \STATE  Calculate volume $\bm{v}_{j}$ and diameter of the mesh through geodesic distance $\bm{g}_{j}$ for each 3D model in $\mathcal{S}$.
    \ENDFOR
    \STATE  Arrange all the $n$ histograms $\bm{h}$ into a $n\times k$ data matrix $\bm{X}=(\bm{x}_1,\dots,\bm{x}_n)^{T}$.
    \STATE  Aggregate mesh diameter, volume $\bm{v}$ and weight $\bm{w}$ to $n\times (k+3)$ data matrix $\bm{X}$.
    \STATE  Perform partial least-squares regression on $\bm{X}$ to find the $n$-dimensional vector $\hat{\bm{y}}$ of predicted cut weights.
    \ENSURE $n$-dimensional vector $\hat{\bm{y}}$ containing predicted weights of pork composition.
  \end{algorithmic}
\end{algorithm}

 We carried out the experiments on a laptop using an Intel Core i$7$ processor with $2.00$ GHz and $16$ GB RAM. Also, our implementation was done in MATLAB. We also considered $301$ eigenvalues and corresponding eigenvectors of the LBO. 
 In this study, we set the resolution parameter as $R = 2$, leading to an SGWS matrix of size $5\times m$, where $m$ is the number of points in our 3D half-carcass model. Besides, we considered four components for our PLS regression. It is noteworthy that the training process is performed offline on concatenated $p\times mn$ SGWS matrices from $n$ meshes in dataset $\mathcal{S}$ achieved by applying k-means algorithm for the dictionary building process.

\begin{table}[b]
\caption{Descriptive characteristics and predicted weight  of primal cuts.}
  \label{Table:primal cuts}
  \centering
  \resizebox{12cm}{!}{
\begin{tabular}{llllllll}
\toprule

Dependent variables\\ $n = 195$ & \quad Mean (kg) &\quad S. D. & \quad Min & \quad Max & \quad $R^{2}$ & \quad RMSE & \quad CVe (\%)\\

\midrule
Ham & \quad $12.032$ & \quad $0.844$  & \quad $9.371$ & \quad $14.003$ & \quad $0.80$  & \quad $0.377$ & \quad $3.13$\\
 \midrule
Shoulder & \quad $12.507$ & \quad $0.879$  & \quad $9.976$ & \quad $14.778$ & \quad $0.79$  & \quad $0.399$ & \quad $3.19$\\
\midrule
Loin & \quad $12.318$ & \quad $0.984$  & \quad $9.448$ & \quad $15.041$ & \quad $0.73$  & \quad $0.507$ & \quad $4.12$\\
\midrule
Belly  & \quad $8.692$ & \quad $0.863$  & \quad $6.288$ & \quad $10.986$ & \quad $0.78$  & \quad $0.406$ & \quad $4.67$\\

\bottomrule
\end{tabular}}
\end{table}

\section{Results and Discussion}\label{experiment}

We assessed the performance of our proposed SpectralWeight framework for measuring hog carcass quality via extensive experiments. We created 3D models of 195 half-carcasses using a 3D scanner, followed by downsampling the mesh surfaces to have roughly $3000$ vertices for each model. We subsequently applied SpectralWeight to extract geometric features of 3D models and then employed the PLS regression to find the best parameters for the weight prediction of pork compositions. The basic idea behind PLS \cite{Wold:84} is to project high-dimensional features into a subspace with a lower dimension. The features in the new subspace, so-called latent features, are a linear combination of the original features. PLS is useful in cases where the number of variables $(k+3)$ in a data matrix $\bm{X}$ are substantially greater than the number of observations $n$. We took advantage of PLS regression since multiple linear regression fails due to multicollinearity among $\bm{X}$ variables. The regression is consequently performed on the latent variables.

\begin{table}[t]
\caption{Descriptive characteristics and predicted weight  of commercial cuts.}
  \label{Table:commercial cuts}
  \centering
  \resizebox{12cm}{!}{
\begin{tabular}{llllllll}
\toprule

Dependent variables\\ $n = 195$ & \quad Mean (kg) &\quad S. D. & \quad Min & \quad Max & \quad $R^{2}$ & \quad RMSE & \quad CVe (\%)\\

\midrule
Pork leg C100 & \quad $11.314$ & \quad $0.816$  & \quad $8.745$ & \quad $13.317$ & \quad $0.80$  & \quad $0.365$ & \quad $3.22$\\
 \midrule
Shoulder picnic  C311 & \quad $4.510$ & \quad $0.446$  & \quad $3.529$ & \quad $5.789$ & \quad $0.54$  & \quad $0.304$ & \quad $6.73$\\
\midrule
Shoulder blade C320 & \quad $4.912$ & \quad $0.457$  & \quad $3.695$ & \quad $5.974$ & \quad $0.64$  & \quad $0.274$ & \quad $5.57$\\
\midrule
Shoulder blade boneless C325  & \quad $4.068$ & \quad $0.380$  & \quad $3.088$ & \quad $5.034$ & \quad $0.60$  & \quad $0.240$ & \quad $5.90$\\
\midrule
Loin C200  & \quad $10.065$ & \quad $0.820$  & \quad $6.849$ & \quad $12.034$ & \quad $0.70$  & \quad $0.446$ & \quad $4.43$\\
\midrule
Loin boneless C201  & \quad $7.370$ & \quad $0.661$  & \quad $4.686$ & \quad $9.392$ & \quad $0.64$  & \quad $0.395$ & \quad $5.36$\\
\midrule
Loin back ribs C505 & \quad $0.675$ & \quad $0.076$  & \quad $0.501$ & \quad $0.914$ & \quad $0.42$  & \quad $0.057$ & \quad $8.52$\\
\midrule
Tenderloin (skinless) C228 & \quad $0.477$ & \quad $0.065$  & \quad $0.262$ & \quad $0.669$ & \quad $0.56$  & \quad $0.043$ & \quad $8.97$\\
\midrule
Belly C400  & \quad $4.636$ & \quad $0.596$  & \quad $3.136$ & \quad $6.369$ & \quad $0.70$  & \quad $0.324$ & \quad $6.98$\\
\midrule
Belly trimmings  & \quad $1.805$ & \quad $0.227$  & \quad $1.187$ & \quad $2.425$ & \quad $0.54$  & \quad $0.154$ & \quad $8.53$\\
\midrule
Side ribs (regular trim) C500  & \quad $1.855$ & \quad $0.214$  & \quad $1.226$ & \quad $2.434$ & \quad $0.51$  & \quad $0.150$ & \quad $8.06$\\
\midrule
Hock C355  & \quad $1.050$ & \quad $0.114$  & \quad $0.606$ & \quad $1.363$ & \quad $0.41$  & \quad $0.088$ & \quad $8.35$\\
\bottomrule
\end{tabular}}
\end{table}

\begin{table}[t]
\caption{Descriptive characteristics and predicted weight  of tissue composition in major commercial cuts.}
  \label{Table:major commercial cuts}
  \centering
  \resizebox{12cm}{!}{
\begin{tabular}{lllllllll}
\toprule

Items\\ $n = 195$ & \quad Tissue composition & \quad Mean (kg) &\quad S. D. & \quad Min & \quad Max & \quad $R^{2}$ & \quad RMSE & \quad CVe (\%)\\
 \cmidrule(r){2-9}

    \multirow{3}{*}{Pork leg C100} & \quad Muscle & \quad $8.014$  & \quad $0.704$ & \quad $5.860$ & \quad $10.124$  & \quad $0.68$ & \quad $0.396$ & \quad $4.95$\\
    &
    \quad Fat & \quad $1.934$  & \quad $0.427$ & \quad $1.115$ & \quad $3.102$  & \quad $0.67$ & \quad $0.243$ & \quad $12.59$\\
    &  
    \quad Bone & \quad $0.941$  & \quad $0.080$ & \quad $0.757$ & \quad $1.172$  & \quad $0.56$ & \quad $0.053$ & \quad $5.62$\\
    &  
    \quad Skin & \quad $0.391$  & \quad $0.061$ & \quad $0.257$ & \quad $0.619$  & \quad $0.28$ & \quad $0.052$ & \quad $13.22$\\

 \cmidrule(r){2-9}

    \multirow{3}{*}{Shoulder picnic C311} & \quad Muscle & \quad $3.016$  & \quad $0.390$ & \quad $1.965$ & \quad $4.053$  & \quad $0.42$ & \quad $0.297$ & \quad $9.83$\\
    &
    \quad Fat & \quad $0.937$  & \quad $0.200$ & \quad $0.484$ & \quad $1.572$  & \quad $0.56$ & \quad $0.133$ & \quad $14.18$\\
    &  
    \quad Bone & \quad $0.380$  & \quad $0.042$ & \quad $0.313$ & \quad $0.563$  & \quad $0.38$ & \quad $0.033$ & \quad $8.66$\\
    &  
    \quad Skin & \quad $0.164$  & \quad $0.024$ & \quad $0.100$ & \quad $0.230$  & \quad $0.23$ & \quad $0.021$ & \quad $12.93$\\
    
 \cmidrule(r){2-9}

    \multirow{3}{*}{Shoulder blade boneless C325} & \quad Muscle & \quad $3.178$  & \quad $0.315$ & \quad $2.263$ & \quad $4.026$  & \quad $0.57$ & \quad $0.207$ & \quad $6.51$\\
    &
    \quad Fat & \quad $0.890$  & \quad $0.169$ & \quad $0.472$ & \quad $1.372$  & \quad $0.53$ & \quad $0.116$ & \quad $13.01$\\
    
 \cmidrule(r){2-9}

    \multirow{3}{*}{Loin boneless C201} & \quad Muscle & \quad $5.847$  & \quad $0.645$ & \quad $3.503$ & \quad $7.516$  & \quad $0.61$ & \quad $0.404$ & \quad $6.91$\\
    &
    \quad Fat & \quad $1.523$  & \quad $0.255$ & \quad $0.757$ & \quad $2.205$  & \quad $0.63$ & \quad $0.155$ & \quad $10.17$\\

 \cmidrule(r){2-9}

    \multirow{3}{*}{Belly C400} & \quad Muscle & \quad $2.689$  & \quad $0.374$ & \quad $1.835$ & \quad $3.620$  & \quad $0.62$ & \quad $0.229$ & \quad $8.53$\\
    &
    \quad Fat & \quad $1.947$  & \quad $0.474$ & \quad $0.778$ & \quad $3.077$  & \quad $0.70$ & \quad $0.260$ & \quad $13.36$\\

 \cmidrule(r){2-9}

    \multirow{3}{*}{Belly Trimmings} & \quad Muscle & \quad $0.811$  & \quad $0.124$ & \quad $0.297$ & \quad $1.154$  & \quad $0.29$ & \quad $0.104$ & \quad $12.80$\\
    &
    \quad Fat & \quad $0.994$  & \quad $0.204$ & \quad $0.512$ & \quad $1.699$  & \quad $0.45$ & \quad $0.151$ & \quad $15.18$\\

\bottomrule
\end{tabular}}
\end{table}

To evaluate the performance of the SpectralWeight, we utilized some performance measurements such as coefficient of determination ($R^{2}-score$), root mean square error ($RMSE$), and coefficient of variation error ($CVe$). It is worth noting that since $CVe$ considers the information of the average weight of the cut, it is a more reliable and fair evaluation metric. To circumvent overfitting, we carefully performed leave-one-out cross-validation over the pork shapes by randomly sampling a set of training instances from our pork carcass dataset for learning and a separate hold-out set for testing. Tables \ref{Table:primal cuts} to \ref{Table:tissue composition} demonstrate the descriptive characteristics and predicted weight of primal cuts, commercial cuts, tissue composition in major commercial cuts, and tissue composition in half-carcasses, respectively.

Table \ref{Table:primal cuts} shows the accuracy of weight prediction for primal cuts. Also, the standard deviation and mean of each primal cut is computed and considered. As can be seen, the lowest prediction error belongs to Ham cut with $CVe=3.13$, while the highest prediction error corresponds to Belly cut with $CVe=4.67$.

Table \ref{Table:commercial cuts} shows the performance of our algorithm for predicting the weights of commercial cuts. As shown, pork leg $C100$ achieved the highest accuracy of prediction with $CVe=3.22$, while Tenderloin (skinless) $C228$ has the lowest accuracy with $CVe=8.97$.

We extended our experiments to further evaluating the major commercial cuts by predicting their tissue composition. The two major commercial cuts of pork leg $C100$ and shoulder picnic $C311$ consist of four tissues, i.e. muscle, fat, bone and skin. As can be observed from Table \ref{Table:major commercial cuts}, best weight predictions correspond to the muscle tissue of pork leg $C100$ and bone tissue of shoulder picnic $C311$ with a correlation of variation error of $4.95$ and $8.66$, respectively. Shoulder blade boneless $C325$, loin boneless $C201$, belly $C400$ and belly trimmings are the other tissues of the major commercial cuts that are composed of only muscle and fat. As shown in Table \ref{Table:major commercial cuts}, for all the four tissues, muscle tissue gained the highest prediction accuracy with $CVe$ of $6.51$, $6.91$, $8.53$, and $12.80$, respectively.

In a bid to investigate the amount of the total tissue composition of muscle, fat, bone, and skin in the half-carcasses, we present Table \ref{Table:tissue composition}, in which the characteristic information of each tissue for the $195$ half-carcasses is demonstrated separately. More precisely, the amount of muscle is achieved by summing up the major commercial cuts including $C100$, $C311$, $C325$, $C201$, $C400$ and the belly trimmings. For the fat, we took into account the major commercial cuts containing $C100$, $C311$, $C325$, $C201$, $C400$, and in the belly trimmings. To calculate the amount of bone, we considered the sum of bone tissue from the half-carcass except the bone tissue contained in the feet, the hock, and the ribs. Also, the amount of skin is obtained by adding the skins from the half-carcass except the skin on the feet, the hock, and the jowl.

As can be seen, our proposed framework is able to predict the weight of muscle tissue with a lower correlation variation of $4.11$ as well as a higher correlation of determination of $R^{2}=0.77$ than the other tissues, respectively. Since muscle is a more valuable tissue for commercial uses, our results for estimating muscle tissue make our algorithm a potential candidate for replacing the traditional methods of carcass quality assessment.

\begin{table}[t]
\caption{Descriptive characteristics and predicted weight  of tissue composition in half-carcass.}
  \label{Table:tissue composition}
  \centering
  \resizebox{12cm}{!}{
\begin{tabular}{llllllll}
\toprule

Items of composition\\ $n = 195$ & \quad Mean (kg) &\quad S. D. & \quad Min & \quad Max & \quad $R^{2}$ & \quad RMSE & \quad CVe (\%)\\

\midrule
Muscle & \quad $23.553$ & \quad $2.041$  & \quad $16.394$ & \quad $28.451$ & \quad $0.77$  & \quad $0.968$ & \quad $4.11$\\
 \midrule
Fat & \quad $10.575$ & \quad $2.158$  & \quad $5.076$ & \quad $16.152$ & \quad $0.73$  & \quad $0.771$ & \quad $9.37$\\
\midrule
Bone & \quad $2.968$ & \quad $0.257$  & \quad $2.377$ & \quad $3.641$ & \quad $0.68$  & \quad $0.145$ & \quad $4.88$\\
\midrule
Skin  & \quad $1.541$ & \quad $0.173$  & \quad $1.100$ & \quad $2.053$ & \quad $0.33$  & \quad $0.141$ & \quad $9.15$\\

\bottomrule
\end{tabular}}
\end{table}

\section{Conclusions}
In this study, we introduced SpectralWeight to estimate the quality of pork carcasses by weight prediction of pork cuts. We first built the spectral graph wavelet signature for every mesh point locally and then aggregated them as a global feature through the bag-of-geometric-words notion. To further ameliorate the discrimination power of SpectralWeight, we merged information of mesh diameter and volume to our pipeline. As the results show, our proposed method can predict the weight of different cuts and tissues of a pork half-carcass with high accuracy and hence is practical to be employed in the pork industry. 
 \section{Acknowledgements}
This work was supported by Swine Innovation Porc within the Swine Cluster 2: Driving Results Through Innovation research program. Funding is provided by Agriculture and Agri‐Food Canada through the AgriInnovation Program, provincial producer organizations and industry partners.  


%
%
 \bibliographystyle{splncs04}
 \bibliography{biblio}

\begin{thebibliography}{10}
\providecommand{\url}[1]{\texttt{#1}}
\providecommand{\urlprefix}{URL }
\providecommand{\doi}[1]{https://doi.org/#1}

\bibitem{CPI:11}
Canadian Pork Handbook and the Distributor Education Program (DEP) 1st Edition.
  Ottawa, Ontario: Canadian Pork International (2011)

\bibitem{Atasoy:16}
Atasoy, S., Donnelly, I., Pearson, J.: Human brain networks function in
  connectome-specific harmonic waves. Nature Communication  \textbf{7} (2016)

\bibitem{Aubry:11}
Aubry, M., Schlickewei, U., Cremers, D.: The wave kernel signature: A quantum
  mechanical approach to shape analysis. In: Proc. Computational Methods for
  the Innovative Design of Electrical Devices (2011)

\bibitem{Bronstein:11}
Bronstein, A., Bronstein, M., Guibas, L., Ovsjanikov, M.: Shape {G}oogle:
  Geometric words and expressions for invariant shape retrieval. ACM Trans.
  Graphics  \textbf{30}(1) (2011)

\bibitem{Chaudhari:14}
Chaudhari, A., Leahy, R., Wise, B., Lane, N., Badawi, R., Joshi, A.: Global
  point signature for shape analysis of carpal bones. Physics in Medicine and
  Biology  \textbf{59},  961--973 (2014)

\bibitem{Daumas:96}
Daumas, G., Dhorne, T.: Historique et futur du classement objectif des
  carcasses de porc en france. Journ{\'e}es de la recherche porcine en France
  \textbf{28},  171--180 (1996)

\bibitem{Engel:03}
Engel, B., Buist, W.G., Walstra, P., Olsen, E., Daumas, G.: Accuracy of
  prediction of percentage lean meat and authorization of carcass measurement
  instruments: adverse effects of incorrect sampling of carcasses in pig
  classification. Animal Science  \textbf{76},  199--209 (2003)

\bibitem{Fredeen:68}
Fredeen, H.: Pig breeding in {C}anada. World Review of Animal Production
  \textbf{2},  87--95 (1968)

\bibitem{Gao:14}
Gao, Z., Yu, Z., Pang, X.: A compact shape descriptor for triangular surface
  meshes. Computer-Aided Design  \textbf{53},  62--69 (2014)

\bibitem{Gispert:07}
Gispert, M., Furnols, F., Gil, M., Velarde, A., Diestre, A., Carrión, D.,
  Sosnicki, A., Plastow, G.: Relationships between carcass quality parameters
  and genetic types. Meat Science  \textbf{77}(3),  397--404 (2007)

\bibitem{kimmel:98}
Kimmel, R., Sethian, J.A.: Computing geodesic paths on manifolds. In Proc. of
  the national academy of Sciences  \textbf{95}(15),  8431--8435 (1998)

\bibitem{Levy:06}
L\'evy, B.: Laplace-{B}eltrami eigenfunctions: Towards an algorithm that
  ``understands'' geometry. In: SMI (2006)

\bibitem{Marcoux:05}
Marcoux, M., Faucitano, L., Pomar, C.: The accuracy of predicting carcass
  composition of three different pig genetic lines by dual-energy x-ray
  absorptiometry. Meat Science  \textbf{70},  655--663 (2005)

\bibitem{Masoumi:17}
Masoumi, M., {Ben Hamza}, A.: Spectral shape classification: A deep learning
  approach. Journal of Visual Communication and Image Representation
  \textbf{43} (2017)

\bibitem{Masoumi:16}
Masoumi, M., Li, C., {Ben Hamza}, A.: A spectral graph wavelet approach for
  nonrigid 3{D} shape retrieval. Pattern Recognition Letters  \textbf{83}
  (2016)

\bibitem{Masoumi:18b}
Masoumi, M., Rezaei, M., Hamza, A.B.: Global spectral graph wavelet signature
  for surface analysis of carpal bones. Physics in Medicine \& Biology
  \textbf{63}(3) (2018)

\bibitem{Masoumi:19a}
Masoumi, M., Toews, M., Lombaert, H.: Waveletbrain: Characterization of human
  brain viaspectral graph wavelets. arXiv:1906.06158v2  (2019)

\bibitem{Melzi:19}
Melzi, S., Ren, J., Rodol{\`a}, E., Sharma, A., Wonka, P., Ovsjanikov, M.:
  Zoomout: Spectral upsampling for efficient shape correspondence. ACM
  Transactions on Graphics (TOG)  \textbf{38}(6), ~155 (2019)

\bibitem{Meyer:03}
Meyer, M., Desbrun, M., Schr\"oder, P., Barr, A.: Discrete
  differential-geometry operators for triangulated 2-manifolds. Visualization
  and mathematics III  \textbf{3}(7) (2003)

\bibitem{Nissen:06}
Nissen, P., Busk, H., Oksama, M., Seynaeve, M., Gispert, M., Walstra, P.,
  Hansson, I., Olsen, E.: The estimated accuracy of the {EU} reference
  dissection method for pig carcass classification. Meat Science
  \textbf{73}(1),  22--28 (2006)

\bibitem{Picouet:10}
Picouet, P., Teran, F., Gispert, M., i~Furnols, M.F.: Lean content prediction
  in pig carcasses, loin and ham by computed tomography ({CT}) using a density
  model. Meat Science  \textbf{86},  616–622 (2010)

\bibitem{Pomar:03}
Pomar, C., Marcoux, M.: Comparing the {C}anadian pork lean yields and grading
  indexes predicted from grading methods based on {D}estron and {H}ennessy
  probe measurements. Canadian Journal of Animal Science  \textbf{83},
  451--458 (2003)

\bibitem{Pomar:09}
Pomar, C., Marcoux, M., Gispert, M., i~Furnols, M.F., Daumas, G.: Determining
  the lean content of pork carcasses. Improving the sensory and nutritional
  quality of fresh meat pp. 493--518 (2009)

\bibitem{qiao:19}
Qiao, Y., Gao, L., Yang, J., Rosin, P.L., Lai, Y., Chen, X.: Laplacian{N}et:
  Learning on 3{D} meshes with laplacian encoding and pooling (2019)

\bibitem{Reuter:06}
Reuter, M., Wolter, F., Peinecke, N.: Laplace-{B}eltrami spectra as
  '{S}hape-{DNA}' of surfaces and solids. Computer-Aided Design  \textbf{38}(4)
  (2006)

\bibitem{Rodola:SHREC17}
Rodola, E., Cosmo, L., O.Litany, Bronstein, M.M., Bronstein, A.M., Audebert,
  N., Hamza, A.B., Boulch, A., Castellani, U., Do, M.N., Duong, A.D., Furuya,
  T., Gasparetto, A., Hong, Y., Kim, J., Saux, B.L., Litman, R., {Masoumi}, M.,
  Minello, G., Nguyen, H.D., Nguyen, V.T., Ohbuchi, R., Pham, V.K., Phan, T.V.,
  Rezaei, M., Torsello, A., Tran, M.T., Tran, Q.T., Truong, B., Wan, L., Zou11,
  C.: {SHREC}'17 track: Deformable shape retrieval with missing parts. In:
  Proc. Eurographics Workshop on 3D Object Retrieval 2017. pp.~1--9 (2017)

\bibitem{Rustamov:07}
Rustamov, R.: Laplace-{B}eltrami eigenfunctions for deformation invariant shape
  representation. In: Proc. Symp. Geometry Processing. pp. 225--233 (2007)

\bibitem{Shirley:07}
Shirley, A.: Determination of fat, moisture, and protein in meat and meat
  products by using the {FOSS} {F}oodscan near-infrared spectrophotometer with
  {FOSS} artificial neural network calibration model and associated database:
  collaborative study. Journal of AOAC International  \textbf{90}(4),
  1073--1083 (2007)

\bibitem{Sun:09}
Sun, J., Ovsjanikov, M., Guibas, L.: A concise and provably informative
  multi-scale signature based on heat diffusion. Computer Graphics Forum
  \textbf{28}(5) (2009)

\bibitem{Vester-Christensen:09}
Vester-Christensen, M., Erbou, S., Hansen, M., Olsen, E., Christensen, L.,
  Hviid, M., Ersbøll, B., Larsen, R.: Virtual dissection of pig carcasses.
  Meat Science  \textbf{81},  699--704 (2009)

\bibitem{Wang:20}
Wang, Y., Ren, J., Yan, D., Guo, J., Zhang, X., Wonka, P.: {MGCN}: Descriptor
  learning using multiscale {GCN}s. arXiv preprint arXiv:2001.10472  (2020)

\bibitem{Wold:84}
Wold, S., Ruhe, A., Wold, H., Dunn, W.J.: The collinearity problem in linear
  regression. {T}he partial least squares (pls) approach to generalized
  inverses. SIAM Journal on Scientific and Statistical Computing  \textbf{5}(3)
  (1984)

\bibitem{YI:17}
Yi, L., Su, H., Guo, X., Guibas, L.: {SyncSpecCNN}: Synchronized spectral {CNN}
  for 3{D} shape segmentation. In: CVPR. pp. 2282--2290 (2017)

\end{thebibliography}


%
%
%
%
%
\end{document}